\documentstyle[12pt]{article}
\begin{document}
   
\author{ Fevzi B\"{U}Y\"{U}KKILI\c{C} , Do\v{g}an DEM\.{I}RHAN , 
U\v{g}ur TIRNAKLI \thanks{e-mail: tirnakli@fenfak.ege.edu.tr}\\
Department of Physics, Faculty of Science, Ege University \\
35100, Bornova \.{I}zmir-TURKEY}

\title{Generalization of the Mean-Field Ising Model Within Tsallis
Thermostatistics}

\date{ }

\maketitle
   
\begin{abstract}
In this study, the mean-field Ising model, using the Bogolyubov inequality
which has been obtained in the framework of the generalized statistical
thermodynamics (GST), and is suitable for nonextensive systems, has been
investigated. Generalized expressions for the mean-field magnetization and
free energy have been established. These new results have been verified by
the fact that they transform to the well-known Boltzmann-Gibbs results in
the $q\rightarrow 1$ limiting case. For the index $q$ which characterizes
the fractal structure of the magnetic system, an interval has been
established where the generalized mean-field free energy has a minimum and
mean-field magnetization has a corresponding finite value. The interval of
$q$ is consistent with paramagnetic free spin systems [14].

\end{abstract}

\section{Introduction}

Boltzmann-Gibbs statistics is used in Physics to study the systems having
the conditions (i) the spatial range of the microscopic interactions are
short-ranged (ii) the time range of the microscopic memory is short-ranged
(iii) the system evolves in a Euclidean-like space-time. These kind of
systems are said to be extensive. Whenever a system does not obey these
restrictions (non-extensive systems), Boltzmann-Gibbs statistics fails and a
non-extensive formalism of statistics must be needed.

In the recent years, a trend towards the non-extensive physics is rapidly
increasing. In this context, the endeavour of the generalization of some of
the conventional concepts such as entropy, free energy, etc. have been under
investigation. These generalizations could roughly be classified under two
subject titles: namely the generalized statistical thermodynamics (GST) and
quantum groups. In a recent paper [1], the author called attention to a
possible relation between these two subjects which appear to be very
distinct.

In 1988, C. Tsallis proposed a generalized entropy [2]

\begin{equation}
S_q=k\frac{1-\sum_{i=1}^{W}p_i^q}{q-1}
\end{equation}

\noindent where $k$ is a positive constant, $p_i$ is the probability of the
system to be in a microstate, $W$ is the total number of the configurations
of the system and $q$ is a parameter relevant to the measure. Tsallis
entropy has the usual standard entropy properties such as positivity,
equiprobability, concavity and irreversibility but in general lacks the
property of additivity. This entropy transforms to the well-known additive
Shannon entropy in $q\rightarrow 1$ limit.

From the year 1988 up to the present days, numerous concepts of the
statistical mechanics and thermodynamics have been tackled in the frame of
this generalization. Amongst them, the specific heat of the harmonic
oscillator [3], one-dimensional Ising model [4], the Boltzmann H-theorem
[5], the Ehrenfest theorem [6], the von Neumann equation [7], quantum
statistics [8,9], the fluctuation-dissipation theorem [10], Langevin and
Focker-Planck equations [11], paramagnetic systems [14], Callen identity
[15], infinite-range spin-$1/2$ Ising ferromagnet [16], generalization of
Planck radiation law [17,18], Kolmogorov entropy and fractals [19], quantum
uncertainty [20], anisotropic rigid rotator [21], Haldane exclusion
statistics [22,23] and Bose-Einstein condensation [24] could be enumerated.

The generalization which has been successfully applied to the preceding
concepts has also been prosperous in some of the physical applications where
the Boltzmann-Gibbs statistics was inadequate. The establishment of finite
mass for the astrophysical systems in the frame of polytropic structures
[25], the calculation of the specific heat of the unionized hydrogen atom
[26], the derivation of Levy-type anomalous diffusion [27-30], the
construction of a comprehensive thermodynamic description of $d=2$ Euler
turbulence [31] and solar neutrino problem [32] are the examples that could
be mentioned herewith. For a review of the subject see [33] and the
mathematical investigations of the $q$-entropies are now available in
[34-36].

In this study, starting from the generalized Bogolyubov inequality [12], the
generalized free energy is found for the mean-field Ising model. In addition
to this, within GST, the generalized magnetization of the mean-field Ising
model is obtained and the critical temperature is found to be the same as
the expression given by Sarmento in ref.[15]. The graphical solutions of the
generalized magnetization are examined and it is observed that the finite
solutions relevant to the corresponding generalized free energy exist only
for a range of $q$ values.

\section{Ising Model and the Bogolyubov Inequality}

One of the successful models of the interacting spin systems is the Ising
model. A spin variable $s_i$, which is allowed to take the values $\mp 1$,
is placed on each lattice site. The spins interact according to the
Hamiltonian

\begin{equation}
{\cal H}=-J\sum_{<ij>}s_i s_j-\lambda\sum_{i}s_i
\end{equation}

\noindent where $J$ is the exchange energy and the first term is responsible
for the ferromagnetism, i.e. the cooperative phenomenon and the possibility
of a phase transition. $\left<ij\right>$ denotes a sum over nearest neighbour 
spins. Here the most important interaction is the exchange interaction $J$ 
and is very dependent on interparticle distance. Therefore the nearest 
neighbours interactions have been taken into account and distant interactions 
ignored. If $J=0$ the eq.(2) is the Hamiltonian of  paramagnet. $\lambda$ is 
termed as the mean-field having the dimension of the magnetic field and could 
be considered as a variation parameter. Unfortunately, very few statistical
models have been exactly solved. In three dimensions not even the nearest
neighbour spin model is tractable. Therefore it is necessary to resort to
approximation methods. One of the most widely used amongst them is
P.R.Weiss' mean-field theory approach [37].

For a given Hamiltonian one way of deriving the mean-field is to start from
the Bogolyubov inequality

\begin{equation}
F\leq\Phi =F_{0}+\left<{\cal H}-{\cal H}_{0}\right>_{0}
\end{equation}

\noindent where ${\cal H}$ is the actual Hamiltonian under consideration but 
not easily handled for the solution and ${\cal H}_{0}$ is a trial Hamiltonian 
possesing a solution which depends on the parameter $\lambda$. $F$ and $F_0$ 
are respectively the free energies associated with the Hamiltonians ${\cal
H}$ and ${\cal H}_0$. $\left<\cdots\right>_{0}$ denotes an average taken in 
the ensemble defined by ${\cal H}_0$.

In order to calculate the right hand side of eq.(3) the trial Hamiltonian
${\cal H}_0$ has been chosen such that; each variable in the system interact 
not with the other variable but with an effective field. This interaction could
be defined in terms of a parameter $\lambda$. On the other hand, the
mean-field is described as the minimum value of eq.(3) with respect to
$\lambda$:

\begin{equation}
F_{mf}=min_{\lambda}\{\Phi\}
\end{equation}

We consider the nearest neighbour Ising model in a zero field, defined by
the Hamiltonian given in eq.(2) with $\lambda =0$, on a lattice of $N$ sites
with each site having $z$ nearest neighbours (coordination number, e.g. for
the simple cubic lattice $z=0$). Thus it follows that 

\begin{equation}
{\cal H}=-J\sum_{<ij>}s_i s_j \;\;\; .
\end{equation}

\noindent The trial Hamiltonian which decomposes into a sum of functions of
individual spins and satisfies the requirement of translational invariance
could be expressed in terms of a single parameter $\lambda$

\begin{equation}
{\cal H}_{0}=-\lambda\sum_{i}s_i \;\;\; .
\end{equation}

\noindent ${\cal H}_0$ is the Hamiltonian of a paramagnet. In this case the
difference of the Hamiltonians is expressed by

\begin{equation}
{\cal H}-{\cal H}_{0}=-J\sum_{<ij>}s_i s_j+\lambda\sum_{i}s_i \;\;\; .
\end{equation}

\section{Bogolyubov Inequality in the Generalized Statistical Thermodynamics
(GST)}

A.Plastino and C.Tsallis have obtained the Bogolyubov inequality in GST with
the variational methods as a result of the concavity (not necessarily
extensivity) property of the Tsallis entropy [12] :

\begin{eqnarray}
\begin{array}{ll}
F\leq\frac{F_0}{H}+\left(1-\frac{1}{H}\right)\frac{1}{\beta
(1-q)}\;\;\;\;\;\;\; & \mbox{for $q<2$}
\end{array}
\end{eqnarray}

\begin{eqnarray}
\begin{array}{ll}
\;\;\;\;=\frac{F_0}{H}-\left(1-\frac{1}{H}\right)\frac{1}{\beta}
\;\;\;\;\;\;\;\;\;\;\;\;\;\; & \mbox{for $q=2$}
\end{array}
\end{eqnarray}
     
\begin{eqnarray}
\begin{array}{ll}
\;\;\;\;\geq\frac{F_0}{H}+\left(1-\frac{1}{H}\right)\frac{1}{\beta
(1-q)}\;\;\;\;\;\;\; & \mbox{for $q>2$}
\end{array}
\end{eqnarray}

\noindent where

\begin{equation}
H\equiv\left<\frac{1-\beta (1-q){\cal H}_0}{1-\beta (1-q){\cal H}}\right>_0
 \;\;\; ,
\end{equation}

\noindent $\beta=1/k_BT$, $k_B$ is the Boltzmann constant and T is the
temperature of the system.

In the $q\rightarrow 1$ limit one gets

\begin{equation}
H\approx 1+\beta (1-q)\left<{\cal H}-{\cal H}_0\right>_0
\end{equation}

\noindent and in this manner eqs.(8),(9),(10) transform to the original
Bogolyubov inequality given by eq.(3).

The partition function belonging to the Ising model corresponding to the
Hamiltinian ${\cal H}_0$, for an individual spin, could be written in GST 
with the expression

\begin{equation}
\left(Z_0\right)_q=\sum_{i}\left[1-\beta
(1-q)\left({\cal H}_0\right)_i\right]^{\frac{1}{1-q}}\;\;\;\; ,
\end{equation}

\noindent namely

\begin{equation}
\left(Z_0\right)_q=\left[1+\beta\lambda
(1-q)\right]^{\frac{1}{1-q}}+\left[1-\beta\lambda
(1-q)\right]^{\frac{1}{1-q}}\;\;\;\; .
\end{equation}

A major difficulty arises from the nonextensivity (nonadditivity) of GST
when we're dealing with the system which is composed of $N$ non-interacting
single-spins. The difficulty is that the partition function of the system
with $N$ independent spin cannot be factorized into the product of $N$
partition functions for system with single spin as it can be done in
standard formalism $(q=1)$. More explicitly this unusual property arises
from the following inequality :

\begin{equation}
\left[1+(1-q)(A+B)\right]^{\frac{1}{1-q}}\neq\left[1+(1-q)A\right]^
{\frac{1}{1-q}}\left[1+(1-q)B\right]^{\frac{1}{1-q}}\;\;\; .
\end{equation}

\noindent A closed analytical expression for the $N$ independent many-body
system can only be obtained by following an approximation scheme like in
ref.[8], namely by taking eq.(15) as an equality. This problem has also been
pointed out by Pennini et al. [38] for the generalized distribution
functions and they made a numerical application for a simple system without
an approximation (see also [39]). In addition to this, eq.(15) has widely
analyzed in ref.[18] and it is realized that this factorization scheme
provides a bound to the exact answer. In fact, it can be seen that this
procedure is analogous to the standard mean-field approach, which, through
Bogolyubov inequality, also is a bound of the exact result.

Therefore, by using this approximation scheme, in GST the free energy of a
system which is composed of $N$ non-interacting single-spins becomes

\begin{equation}
\left(F_0\right)_q=\frac{\left\{\left[1+(1-q)\beta\lambda\right]^{\frac{1}{1-q}}
+\left[1-(1-q)\beta\lambda\right]^{\frac{1}{1-q}}\right\}^{N(1-q)}-1}
{\beta (1-q)}\;\;\; .
\end{equation}

\noindent On the other hand, in the $q\rightarrow 1$ limit eq.(16)
transforms to the standard expression

\begin{equation}
\left(F_0\right)_1=-\frac{N}{\beta}\log\cosh (\beta\lambda )\;\;\; .
\end{equation}

Within our factorization scheme,generalized average value of a single spin
within a system of $N$ independent spins is assumed to be equal to the mean
value corresponding to a system consisting of just one spin. Since the
average individual spin is $\left(\left<s\right>_0\right)_q=\sum_{i}s p_i^q$,
in GST one gets

\begin{equation}
\left(\left<s\right>_0\right)_q=\frac{\left[1+\beta\lambda (1-q)\right]^
{\frac{q}{1-q}}-\left[1-\beta\lambda (1-q)\right]^{\frac{q}{1-q}}}
{\left[1+\beta\lambda (1-q)\right]^
{\frac{q}{1-q}}+\left[1-\beta\lambda (1-q)\right]^{\frac{q}{1-q}}}\;\;\;\; . 
\end{equation}

\noindent Similarly, in the $q\rightarrow 1$ limit eq.(18) transforms to
the well-known expression

\begin{equation}
\left(\left<s\right>_0\right)_1=\tanh(\beta\lambda )\;\;\;\; .
\end{equation}

\noindent Taking into consideration eqn.(5) and (6), performing the sums
one obtains

\begin{equation}
H=\frac{1+\beta (1-q)\lambda N
\left(\left<s\right>_0\right)_q}{1+\beta (1-q)\frac{1}{2}JzN
\left(\left<s\right>_0^2\right)_q}
\end{equation}

\noindent where the factorization of the interaction term is possible since
$H_0$ includes only the single-site terms. In eq.(20), for a translationally
invariant system $\left(\left<s_i\right>_0\right)_q=
\left(\left<s_j\right>_0\right)_q=
\left(\left<s\right>_0\right)_q$ and $zN/2$ is the number of bonds on the
lattice.

Substitution of eqs.(16) and (20) in the right-hand side of the Bogolyubov
inequality given by eq.(8), leads to the following relation

\begin{equation}
\Phi_q=\frac{\left(F_0\right)_q\left[1+\beta (1-q)\frac{1}{2}JzN
\left(\left<s\right>_0^2\right)_q\right]+\lambda N
\left(\left<s\right>_0\right)_q-\frac{1}{2}JzN
\left(\left<s\right>_0^2\right)_q}
{1+\beta (1-q)\lambda N\left(\left<s\right>_0\right)_q}\;\;\; .
\end{equation}

\noindent In the $q\rightarrow 1$ limit, by taking into consideration
eqs.(17) and (19), one gets from eq.(21) :

\begin{equation}
\Phi_1=\left(F_0\right)_1+\lambda N\left(\left<s\right>_0\right)_1-
\frac{1}{2}JzN\left(\left<s\right>_0^2\right)_1
\end{equation}

\noindent which is identical with the result obtained from eq.(3) with the
same approximation using the Boltzmann distribution [40]. $\Phi_q$ could be
minimized by choosing

\begin{equation}
\lambda =Jz\left(\left<s\right>_0\right)_q\;\;\;\; .
\end{equation}

\noindent This minimum of the mean-field magnetization is substituted in
eq.(18) giving a new expression for the magnetization which could be named
as {\it generalized mean-field magnetization}

\begin{equation}
\left(\left<s\right>_0\right)_q=f\left[\left(\left<s\right>_0\right)_q\right]=
\frac{\left[1+\beta Jz(1-q)\left(\left<s\right>_0\right)_q\right]^
{\frac{q}{1-q}}-
\left[1-\beta Jz(1-q)\left(\left<s\right>_0\right)_q\right]^
{\frac{q}{1-q}}}
{\left[1+\beta Jz(1-q)\left(\left<s\right>_0\right)_q\right]^
{\frac{q}{1-q}}+
\left[1-\beta Jz(1-q)\left(\left<s\right>_0\right)_q\right]^
{\frac{q}{1-q}}}
\end{equation}

\noindent which in the $q\rightarrow 1$ transforms to the expected result

\begin{equation}
\left(\left<s\right>_0\right)_1=\tanh\beta Jz\left(\left<s\right>_0
\right)_1\;\;\; .
\end{equation}

\noindent i.e. identical with the original mean-field magnetization [40].

In order to determine the critical temperature, the gradients of the curves
$y=\left(\left<s\right>_0\right)_q$ and $y=f\left[
\left(\left<s\right>_0\right)_q\right]$ should be equated at the origin
which leads to the result

\begin{equation}
T_c=\frac{qJz}{k}
\end{equation}

\noindent which is identical with the result given in ref.[15]. Standard
value of critical temperature $T_c=Jz/k_B$ is recovered in the $q\rightarrow
1$ limit. It is important to note that $T_c$ depends on $z$ (the number of
nearest neighbours) as well as $q$, but not on the other parameters of the
lattice structure such as dimensionality.

{\it Generalized mean-field free energy} corresponding to the minimum
$\lambda$ is obtained by the substitution of eq.(23) into eq.(21) :

\begin{equation}
\left(F_{mf}\right)_q=\frac{\left(F_0\right)_q\left[1+\beta (1-q)
\frac{1}{2}JzN\left(\left<s\right>_0^2\right)_q\right]+
\frac{1}{2}JzN\left(\left<s\right>_0^2\right)_q}
{1+\beta (1-q)JzN\left(\left<s\right>_0^2\right)_q}\;\;\; .
\end{equation}

\noindent As usual, in the $q\rightarrow 1$ limiting case eq.(27)
transforms to be the conventional result [40]

\begin{equation}
\left(F_{mf}\right)_1=\left(F_0\right)_1+\frac{1}{2}JzN
\left(\left<s\right>_0^2\right)_1\;\;\; .
\end{equation}

\noindent By means of the Bogolyubov inequality given by eqs.(8)-(10) in GST 
and the choice of an easily solvable expression for ${\cal H}_0$ in a
paramagnetic system, the right-hand side of the inequality could be solved
exactly and an upperbound for the free energy of the actual Hamiltonian
could be established. The most probable free energy $\left(F_{mf}\right)_q$
of the system in terms of the free energy $\left(F_0\right)_q$ of the
individual spins are in the form of eq.(27) which has been obtained from
the mean-field theory. Corresponding to eqs.(8),(9) and (10), the
Bogolyubov inequality takes the following forms in GST :

\begin{eqnarray}
\begin{array}{ll}
F\leq\left(F_{mf}\right)_q\;\;\;\;\;\;\; & \mbox{for $q<2$}
\end{array}
\end{eqnarray}

\begin{eqnarray}
\begin{array}{ll}
F=\left(F_{mf}\right)_q\;\;\;\;\;\;\; & \mbox{for $q=2$}
\end{array}
\end{eqnarray}

\begin{eqnarray}
\begin{array}{ll}
F\geq\left(F_{mf}\right)_q\;\;\;\;\;\;\; & \mbox{for $q>2$}
\end{array}
\end{eqnarray}

\section{Graphical Solutions of Generalized Mean-Field Magnetization
Consistent with the Generalized Mean-Field Free Energy}

In the standard mean-field Ising model, the unique solution of the
magnetization for temperatures greater than $T_c$, is 
$\left(\left<s\right>_0\right)_1=0$ which corresponds to the paramagnetic phase.
On the other hand, for temperatures less than $T_c$, there exist two
solutions corresponding to the values $\left(\left<s\right>_0\right)_1=0$ and 
finite $\left(\left<s\right>_0\right)_1$. The mean-field free energy corresponding 
the finite $\left(\left<s\right>_0\right)_1$ value is a minimum and the system 
under consideration is in stable ferromagnetic phase. The minima of the
generalized mean-field free energy for various $q$ values are presented in
Fig.1 whereas the finite values of the corresponding generalized mean-field
magnetization is demonstrated in Fig.2 (standard values appear in the case
of $q=1$).

As it is recognized from Fig.1, the mean-field free energy exhibits a
minimum for $q$ values in the interval $1<q<3$. In this case, it is expected
that, the generalized mean-field magnetization possesses a finite solution.
From the curve plotted in Fig.2, it is noticed that, $y=f\left[\left(
\left<s\right>_0\right)_q\right]$ has a solution only if $q$ takes values in
the interval $1<q<3$. This result coincides with the $q$ values calculated
by another method which is completely different from the approximation
method used in this study where the distributions of the total magnetic
moments of the magnetic systems in the thermodynamical equilibrium have been
taken into account [14].

\section{Conclusions}

In this study, a step has been taken towards the enlargement of the areas of
applications in the generalized statistical thermodynamica (GST). In this
manner, our effort in the aspect of the investigation of the phase
transitions with GST has been strengthened. In this framework, by the
approximate scheme which we have used here, closed and analytical
expressions for the generalized mean-field magnetization and the
generalized mean-field free energy have been derived for the $N$ independent
single-spin systems. From the graphical representations it is seen that, the
stable solutions consistent with each other are obtained if and only if $q$
lies in the interval $1<q<3$. This interval is the same as the one obtained
in ref.[14]. In the $q\rightarrow 1$ case, it is observed that, both of the
two expressions transforms to the results obtained in the standard
Boltzmann-Gibbs statistics. In addition to this, $q$-dependent mean-field
critical temperature $T_c(q)$ is deduced. It is observed that our result
which is obtained for the square lattice within a renormalization group
calculation [39]. The consistency of these results can be taken as a
validification of our approximation.

\section*{Acknowledgments}

It is a great pleasure for the authors to express their sincere thanks to
Professor C. Tsallis (CBPF, Brazil) and Professor T. Altanhan (Ankara
University, Turkey) for their kind interest and bringing to their attention
the present status of the subject. We are indebted to TUBITAK for providing
our communication with Professor C. Tsallis.

\newpage

\section*{Figure Captions}

\vspace{2cm}

\noindent
Fig. 1. Generalized mean-field free energy versus generalized mean-field
magnetization for various $q$ values. Standard result is depicted for $q=1$
case.

\vspace{1cm}

\noindent
Fig. 2. Graphical solution of the generalized mean-field magnetization. The
solid line is the curve $y=\left(\left<s\right>_0\right)_q$, while the
others represent $y=f\left[\left(\left<s\right>_0\right)_q\right]$
(eq.(24)) plots for various $q$ values. Standard result is depicted for
$q=1$ case.


\newpage

\begin{thebibliography}{99}


\bibitem{tsallis1}  C. Tsallis, Phys. Lett. A195 (1994) 329.

\bibitem{tsallis0}  C. Tsallis, J. Stat. Phys. 52 (1988) 479.

\bibitem{ito} N. Ito and C. Tsallis, Nuovo Cimento 11 (1989) 907.

\bibitem{andrade} R.F.S. Andrade, Physica A175 (1991) 285; A203 (1994) 486.

\bibitem{mariz} A.M. Mariz, Phys. Lett. A165 (1992) 409; J.D. Ramshaw, Phys.
Lett. A175 (1993) 169 and 171.

\bibitem{plastino1} A. Plastino and A.R. Plastino, Phys. Lett. A177 (1993)
177.

\bibitem{plastino2} A.R. Plastino and A. Plastino, Physica A202 (1994) 438.

\bibitem{buyukkilic1} F. B\"{u}y\"{u}kk{\i}l{\i}\c{c} and D. Demirhan, Phys.
Lett. A181 (1993) 24; F. B\"{u}y\"{u}kk{\i}l{\i}\c{c}, D. Demirhan and A.
G\"{u}le\c{c}, Phys. Lett. A197 (1995) 209.

\bibitem{curilef1} S. Curilef, Z. Phys. B (1996), in press.

\bibitem{dasilva} E.P. da Silva, C. Tsallis and E.M.F. Curado, Physica A199
(1993) 137; 203 (1994) E160; A. Chame and E.M.L. de Mello, J. Phys. A27
(1994) 3663.

\bibitem{stariolo1} D.A. Stariolo, Phys. Lett. A185 (1994) 262.

\bibitem{plastino3} A. Plastino and C. Tsallis, J. Phys. A26 (1993) L893.

\bibitem{plastino4} A.R. Plastino, A. Plastino and C. Tsallis, J. Phys. A27
(1994) 5707.

\bibitem{buyukkilic2} F. B\"{u}y\"{u}kk{\i}l{\i}\c{c} and D. Demirhan, Z.
Phys. B99 (1995) 137.

\bibitem{sarmento} E.F. Sarmento, Physica A218 (1995) 482.

\bibitem{nobre} F.D. Nobre and C. Tsallis, Physica A213 (1995) 337.

\bibitem{tsallis2} C. Tsallis, F.C. Sa Barreto and E.D. Loh, Phys Rev E52
(1995) 1447. 

\bibitem{tirnakli1} U. T{\i}rnakl{\i}, F. B\"{u}y\"{u}kk{\i}l{\i}\c{c} and 
D. Demirhan, "Generalized Distribution Functions and an Alternative Approach
to Generalized Planck Radiation Law", preprint; C. Tsallis, private
communications.

\bibitem{zanette1} D.H. Zanette, Physica A223 (1996) 87.

\bibitem{plastino5} A. Plastino and M. Portesi, Physica A (1996), in press;
M. Portesi and A. Plastino, Proc.IV International Conference on Squeezed
States and Uncertainty Relations (FICSSUR-95, Taiyvan, China) NASA
Conference Publication Series (Dec.1995), in press.

\bibitem{curilef2} S. Curilef and C. Tsallis, Physica A215 (1995) 542.

\bibitem{rajagopal1} A.K. Rajagopal, Phys. Rev. Lett. 74 (1995) 1048;
Physica B212 (1995) 309.

\bibitem{kaniadakis1} G. Kaniadakis, A. Lavagno and P. Quarati, "Generalized
Fractional Statistics", preprint.

\bibitem{curilef3} S. Curilef, Phys. Lett. A (1996), in press.

\bibitem{plastino6} A.R. Plastino and A. Plastino, Phys. Lett. A174 (1993)
384.

\bibitem{lucena} L.S. Lucena, L.R. da Silva and C. Tsallis, Phys. Rev. E51
(1995) 6247.

\bibitem{alemany1} P.A. Alemany and D.H. Zanette, Phys. Rev. E49 (1994)
R956.

\bibitem{tsallis3} C. Tsallis, S.V.F. Levy, A.M.C. Souza and R. Maynard, Phys.
Rev. Lett. 75 (1995) 3589.

\bibitem{zanette2} D.H. Zanette and P.A. Alemany, Phys. Rev. Lett. 75 (1995)
366.

\bibitem{tsallis4} C. Tsallis, A.M.C. de Souza and R. Maynard, in "Levy
Flights and Related Phenomena in Physics", Eds. M.F. Shlesinger, U. Frisch
and G.M. Zaslavsky (Springer, Berlin, 1995), page 269.

\bibitem{boghosian} B.M. Boghosian, Phys. Rev. E53 (1996) 4754. 
              
\bibitem{kaniadakis2}  G. Kaniadakis, A. Lavagno and P. Quarati, Phys.
Lett. B369 (1996) 308.

\bibitem{tsallis5} C. Tsallis, in "New Trends in Magnetism, Magnetic
Materials and their Applications", Eds. J.L. Moran-Lopez and J.M. Sanchez
(Plenum Press, New York, 1994), page 451; in "Chaos, Solitons and Fractals",
Ed. G. Marshall (Pergamon Press, 1994), page 539.

\bibitem{raggio1} G.A. Raggio, J. Math. Phys. 36 (1995) 4785.

\bibitem{guerberoff1} G.R. Guerberoff, P.A. Pury and G.A. Raggio, J. Math.
Phys. 37 (1996) 1790.

\bibitem{guerberoff2} G.R. Guerberoff and G.A. Raggio, J. Math. Phys. 37
(1996) 1776.

\bibitem{weiss} P.R. Weiss, Phys. Rev. 74 (1948) 1493.

\bibitem{pennini1} F. Pennini, A. Plastino and A.R. Plastino, Phys. Lett.
A208 (1995) 309; "Nonextensive Thermostatistics, Pauli Principle and the
Structure of the Fermi Surface", Physica A, in press.

\bibitem{cannas1} S. A. Cannas and C. Tsallis, Z. Phys. B100 (1996) 623.

\bibitem{binney} J.J. Binney, N.J. Dowrick, A.J. Fisher and M.E.J. Newman,
"The Theory of Critical Phenomena", (Clarendon Press, Oxford, 1992).
   
\end{thebibliography}
\end{document}